\documentclass[]{aa}
\usepackage{graphicx,amsmath}
\usepackage{txfonts}
\usepackage{lscape}

\newcommand{\abin}{a_{\rm bin}}

\newcommand{\Tbin}{T_{\rm bin}}

\newcommand{\cs}{c_{\rm s}}

\newcommand{\tirr}{T_{\rm irr}}
\newcommand{\tcool}{t_{\rm cool}}
\newcommand{\taueff}{\tau_{\rm eff}}
\newcommand{\tauu}{\tau_{\rm upper}}
\newcommand{\taul}{\tau_{\rm lower}}
\newcommand{\qcool}{{\cal Q}_{\rm cool}}
\newcommand{\rin}{R_{\rm in}}
\newcommand{\rout}{R_{\rm out}}

\newcommand{\rgap}{R_{\rm gap}}

\newcommand{\st}{{\rm St}}
\begin{document}
\title{Thermal structure of circumbinary discs: Circumbinary planets should be icy not rocky}
\titlerunning{Circumbinary planets should be icy not rocky}
\author{Arnaud Pierens 
 \inst{1}
 \and
  Richard P. Nelson\inst{2}}
\institute{ Laboratoire d'astrophysique de Bordeaux, Univ. Bordeaux, CNRS, B18N, all\'ee Geoffroy Saint-Hilaire, 33615 Pessac, France\\
\email{arnaud.pierens@u-bordeaux.fr}
\and
 Astronomy Unit, Queen Mary University of London, Mile End Road, London, E1 4NS, UK}

\abstract{
The process of forming a circumbinary planet is thought to be intimately related to the structure of the nascent circumbinary disc. It has been shown that the structure of a circumbinary disc depends strongly on 3-dimensional effects and on the detailed modelling of the thermodynamics. Here, we employ 3-dimensional hydrodynamical simulations, combined with a proper treatment of the thermal physics using the RADMC-3D radiation transport code, to examine the location of the snow line in circumbinary discs. The models have application to the circumbinary planets that have been discovered in recent years by the Kepler and TESS transit surveys. We find that the snow line is located in a narrow region of the circumbinary disc, close to the inner cavity that is carved out by the central binary, at typical orbital distances of $\sim 1.5-2$~au for the system parameters considered. In this region, previous work has shown that both grain growth and pebble accretion are likely to be inefficient because of the presence of hydrodynamical turbulence. Hence, in situ planet formation interior to the snow line is unlikely to occur and circumbinary planets should preferentially be icy, not rocky.  
}

\maketitle

\section{Introduction}

The discovery of circumbinary planets such as Kepler-16b \citep[e.g.][]{2011Sci...333.1602D} has sparked new interest in how planets can form in highly perturbed environments such as circumbinary discs.   Due to the constraints resulting from gravitational perturbations induced by the central binary, circumbinary discs are ideal laboratories for testing planet formation theories. The different scenarios for planet formation in circumbinary discs are generally based on the assumption that circumbinary planets were either formed in-situ,  at the location where they are currently observed; or at large distances from the binary. In the latter case, an episode of inward migration must be invoked to explain the present orbits of circumbinary planets, which are generally found to orbit close to the zone of dynamical stability defined by \cite{1999AJ....117..621H}. 

Forming circumbinary planets in-situ has been proven to be difficult in the context of standard planetesimal accretion \citep[e.g.][]{2012ApJ...754L..16P}. It has been shown that large planetesimal eccentricities can result from the tidal interaction with the binary or the eccentric circumbinary disc \citep{2008ApJ...681.1599M, 2010ASSL..366..135K}, resulting in collision velocities that are too high to enable planetesimal growth. Growing a massive planetesimal seed in-situ through pebble accretion also appears to be very challenging. The inner disc is prone to a parametric instability that generates hydrodynamical turbulence \citep{2005A&A...432..743P,2014MNRAS.445.2637B}, the main impact of which is to stir up the pebble layer, resulting in inefficient pebble accretion \citep{2020MNRAS.496.2849P,2021MNRAS.508.4806P}. 

Taken together, the above constraints seem to favor the migration scenario, although this scenario has its own drawbacks. In particular, the stopping locations of circumbinary planets migrating in 2D disc models can be a poor match to the locations of some of the known circumbinary planets \citep{2013A&A...556A.134P,2021A&A...645A..68P}, especially for high-eccentricity binaries like Kepler-34 \citep{2012Natur.481..475W}. This arises because a wide and eccentric tidally-induced inner cavity is formed in the circumbinary disc, and the planet itself tends to acquire significant eccentricity, causing its inward migration to stall too far from the central binary. In contrast, recent studies have shown that when the disc is circularized by a planet opening a partial gap \citep{2017MNRAS.465.4735M,2021A&A...645A..68P}, or because of dust feedback onto the gas when the inner regions of the disc become enriched by inwards-drifting dust \citep{2022MNRAS.513.2563C}, a better fit to the observed orbital parameters can be obtained. Compared to 2D discs, \cite{2023A&A...670A.112P} also found a systemic trend for the size and eccentricity of the inner cavity to be smaller in 3D models that include the effects of stellar heating and radiative cooling on the thermal disc structure. 
These latter results underline the importance of considering 3D effects combined with a proper treatment of the disc thermodynamics to obtain realistic structures for circumbinary discs. 

In this context, we present the results of 3D hydrodynamical simulations of circumbinary discs that include irradiation from the central stars and cooling to estimate the location of the snow line in circumbinary discs, with application to several of the systems known to host circumbinary planets. There have been several previous studies that have focused on the position of the snow line. Using a static model of a circumbinary disc that is heated by stellar irradiation and viscous dissipation, \cite{2013ApJ...768L..15C} found that ice lines tend to lie within the zone of instability, and suggested that rocky planets should not form in these systems. The effect of tidal dissipation was examined in globally viscous circumbinary discs by \cite{2015MNRAS.447.1439S} and \cite{2016ApJ...816...94V}, who found that the combination of viscous dissipation and the damping of density waves driven by the binary tends to push the ice line out beyond 3-5~au.  With respect to these previous studies, our work self-consistently computes the density and temperature structure of the disc using 3D hydrodynamical simulations coupled with Monte Carlo radiative transfer calculations. We also examine the influence of localised viscous heating that arises because of the dissipation of the binary-induced hydrodynamical turbulence described above. We work under the assumption that the low ionisation fraction of the circumbinary discs prevents the development of magnetised turbulence within them \citep[e.g.][]{1996ApJ...457..355G}, such that the only source of heating far from the tidally-induced inner cavity is stellar irradiation. We investigate the dependence of the results on the parameters of the central binary, and apply the results to the putative circumbinary discs of several systems known to host  circumbinary planets.

This paper is structured as follows. We describe the numerical model in Sect.~2 and we present the set-up employed for the simulations in Sect.~3.  In Sect.~4, we present the results of our 3D radiation-hydrodynamical simulations. In Sect. 5, we discuss our results and draw conclusions regarding the location of the snow line in circumbinary discs and the implications that this has for the likely composition of circumbinary planets.

\section{Numerical model}
\subsection{3-dimensional hydrodynamical simulations}
\subsubsection{Equations of motion}
We solve the hydrodynamical equations for the conservation of mass,  momentum and internal energy in spherical coordinates $(r,\theta,\phi)$ (radial, polar, azimuthal), with the origin of the frame located at the centre of mass of the binary.  In particular, the energy equation that is incorporated in the code is given by:
 \begin{equation}
  \frac{\partial e}{\partial t}+{\bf \nabla}\cdot(e {\bf v})=-(\gamma-1) e {\bf \nabla}\cdot {\bf v}+\qcool+{\cal Q}_{ \rm bulk}^+ ,
  \label{eq:energy}
 \end{equation}
where  ${\bf v}$ is the velocity,  and $e=\rho c_v T$ is the internal energy per unit volume, with  $\rho$  the density, $c_v$  the specific heat capacity at constant volume, and $T$ the temperature.  In the previous equation,  $\gamma$ is the adiabatic index, which is set to $\gamma=1.4$. 

We consider a cooling scheme where the temperature is relaxed towards a reference temperature $\tirr$ on a cooling timescale $\tcool$. In Eq.  \ref{eq:energy}, the  cooling rate ${\cal Q}_{\rm cool}$ is therefore given by:
\begin{equation}
\qcool=-\rho c_v\frac{T-\tirr}{\tcool}.
\label{eq:qcool}
\end{equation}
Here, we assume $\tirr$ to correspond to the temperature set by  stellar irradiation, augmented by viscous heating if this is included in the model. During the simulations, this temperature is regularly updated using the RADMC-3D Monte Carlo radiative transfer code \citep{2012ascl.soft02015D}, applying the procedure described below in Sect. \ref{sec:radiative}.  The  cooling timescale, $\tcool$,  is calculated at every time step by considering 
the timescale for radiative loss of energy from a Gaussian sphere with a characteristic length scale corresponding to the gas scale height $H$ \citep{2016ApJ...817..102L}. It is given by:
\begin{equation}
\tcool=\frac{\rho c_v H \taueff}{3\sigma_{\rm SB} T^3},
\end{equation}
where $\sigma_{\rm SB}$ is the Stefan-Boltzmann constant and $\taueff$ is the effective optical depth given by:
\begin{equation}
\taueff=\frac{3}{8}\tau+\frac{\sqrt 3}{4}+\frac{1}{4\tau}.
\end{equation}
Following \cite{2016ApJ...817..102L}, the optical depth $\tau$ is calculated as $1/\tau=1/\tauu+1/\taul$ with:
\begin{equation}
\tauu=\int_z^{z_{\rm max}} \rho(z')\kappa(z')dz'
\end{equation}
and
\begin{equation}
\taul=\int_{z_{\rm min}}^z \rho(z')\kappa(z')dz',
\end{equation}
where the opacity $\kappa$ is computed using the Rosseland mean opacity of \cite{2009ApJ...701..620Z}.

 Finally, the heating source term ${\cal Q}_{ \rm bulk}^+$ is provided by artificial viscous heating resulting from shock damping of the density waves excited by the binary.

\subsubsection{Numerical methods}
The simulations presented in this paper were performed using the multifluid version of FARGO3D \citep{2016ApJS..223...11B}.
Computational units are chosen such that the total mass of the binary is $M_\star=1$, the gravitational constant $G=1$, and the radius $R=1$ in the computational domain corresponds to the binary semi-major axis $\abin$. When presenting the simulation results, unless otherwise stated we use the binary orbital period $\Tbin=2\pi\sqrt{\abin^3/GM_\star}$ as the unit of time.

The computational domain in the radial direction extends from $R_{\rm in}=1.13\,\abin$ to $R_{\rm out}=50\,\abin$ and we employ $684$ logarithmically spaced grid cells.   In the azimuthal direction the simulation domain extends from $0$ to $2\pi$ with $384$ uniformly spaced grid cells. In the meridional direction, the simulation domain covers $3.5$ disc pressure scale heights above and below the disc midplane, and we adopt $48$ uniformly spaced grid cells. We note that this resolution allows us to compute the quasi-steady structure of 3D circumbinary disc models for a range of binary parameters, given the computational resources at our disposal. It is not, however, sufficient to directly capture the hydrodynamical turbulence that is excited through the parametric instability in the inner regions of inviscid circumbinary discs \citep[see][]{2021MNRAS.508.4806P}. It is for this reason that we include a viscous heating term in the radiative transfer calculations described below, since we want to estimate how the dissipation of the turbulence affects the position of the snow line.
 
\subsubsection{Boundary conditions}
We adopt an outflow boundary condition at the inner edge to allow mass to accrete onto the binary. Here, the velocity is set to 0 if directed into the computational domain, such that gas can leave the domain but cannot flow into it. The values for the gas density and internal energy in the ghost zones have the same values as in the first active zone. In the context of 2D models, it has been shown that  employing an open boundary together with a location of the inner disc edge $\rin \approx \abin$ leads to a quasi-stationary disc structure after ${\cal O}(10^4)$ binary orbits. Compared to other boundary conditions, adopting an open boundary also appears to be less sensitive to numerical issues \citep{2017A&A...604A.102T}. 

At the outer radial boundary, we employ a wave-killing zone for $R > 0.88$ $\rout$  to avoid wave reflection at the disc outer edge \citep{2006MNRAS.370..529D}. The impact of this wave-killing zone on the disc shape and eccentricity is expected to be small since the disc structure near the wave-killing zone remains close to initial one.

At the meridional boundaries of the three dimensional domain, an outflow boundary condition is also employed, but with the exception that for the  gas density we follow \cite{2016ApJ...833..126B} and maintain vertical stratification by solving the following condition for hydrostatic equilibrium:
\begin{equation}
\frac{1}{\rho}\frac{\partial}{\partial \theta} (\cs^2 \rho)=\frac{v_\phi^2}{\tan \theta}.
\end{equation}
 where $c_s$ is the sound speed and $v_\phi$ the azimuthal velocity.

\subsection{Radiative transfer calculations}
\label{sec:radiative}
\subsubsection{Radiation sources}

\begin{figure}
\centering
\includegraphics[width=\columnwidth]{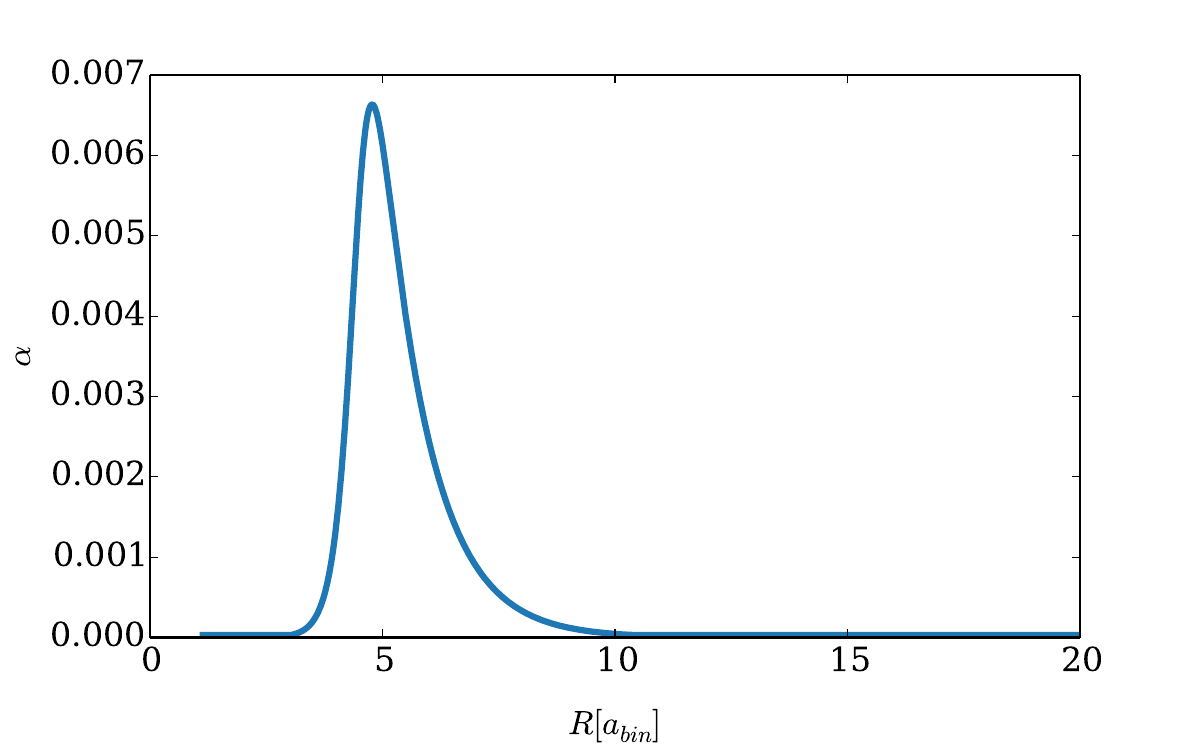}
\caption{$\alpha$ viscous parameter as a function of distance from the binary. The dependence of $\alpha$ on radius is chosen so as to reproduce the $\alpha$ profile found in \cite{2021MNRAS.508.4806P} (see their Fig. 4) where hydrodynamical turbulence operates. }
\label{fig:alpha}
\end{figure}

The radiation  transfer calculations are performed using RADMC-3D  \citep{2012ascl.soft02015D}, in which radiation from  multiple stars can be incorporated, as well as an extra internal  source of energy such as  viscous heating. We assume both stars are located at the centre of the domain and do not take into account the time dependent effects arising from their orbital motion. Regarding the emission from the stars, we first consider their black-body radiation fields using $N=10^8$ photons. In some models, we also take into account localised viscous heating driven by turbulent viscosity, using the modified random walk method \citep{2009A&A...497..155M,2010A&A...520A..70R}. In this case, we employed a reduced number of photons $N=10^7$ due the higher computational cost required to run these models.  The corresponding heat source that is employed in RADMC-3D is given by: 
\begin{equation}
\Gamma_v=\frac{9}{4}\frac{GM_\star}{R^3} \rho \nu,
\end{equation}
where the kinematic viscosity is given by $\nu=\alpha H^2\Omega$ in the context of the $\alpha$ model \citep{1973A&A....24..337S}. Previous work has shown that the most plausible source of turbulence in the inner regions of circumbinary discs is a parametric instability driven by the resonant interaction between inertial-gravity waves and an eccentric mode in the disc \citep{2005A&A...432..743P,2014MNRAS.445.2637B}.  Because it is directly related to the disc eccentricity, turbulence generated in this way is more vigorous close to the inner cavity where the disc eccentricity is the highest, and operates much less efficiently when moving away from the inner cavity. A direct consequence is that the $\alpha$ parameter is not constant, with a peak value of $\alpha \approx 8\times 10^{-3}$ at the cavity edge and $\alpha\approx 2\times 10^{-3}$ at a distance $R\sim 6$ $\abin$. In this work, we adopt the following functional dependence of $\alpha$ on radius:
\begin{equation}
\alpha(R)=0.01\frac{\exp(\rgap-R)}{1+\exp\left(-\frac{R-\rgap}{0.05\rgap}\right)}, 
\end{equation}
where $R_{\rm gap}$ is the estimated gap size (see below). This form is chosen to reproduce the $\alpha$ profile found in \cite{2021MNRAS.508.4806P} (see their Fig. 4), and is plotted in Fig.~\ref{fig:alpha}.

\begin{figure}
\centering
\includegraphics[width=\columnwidth]{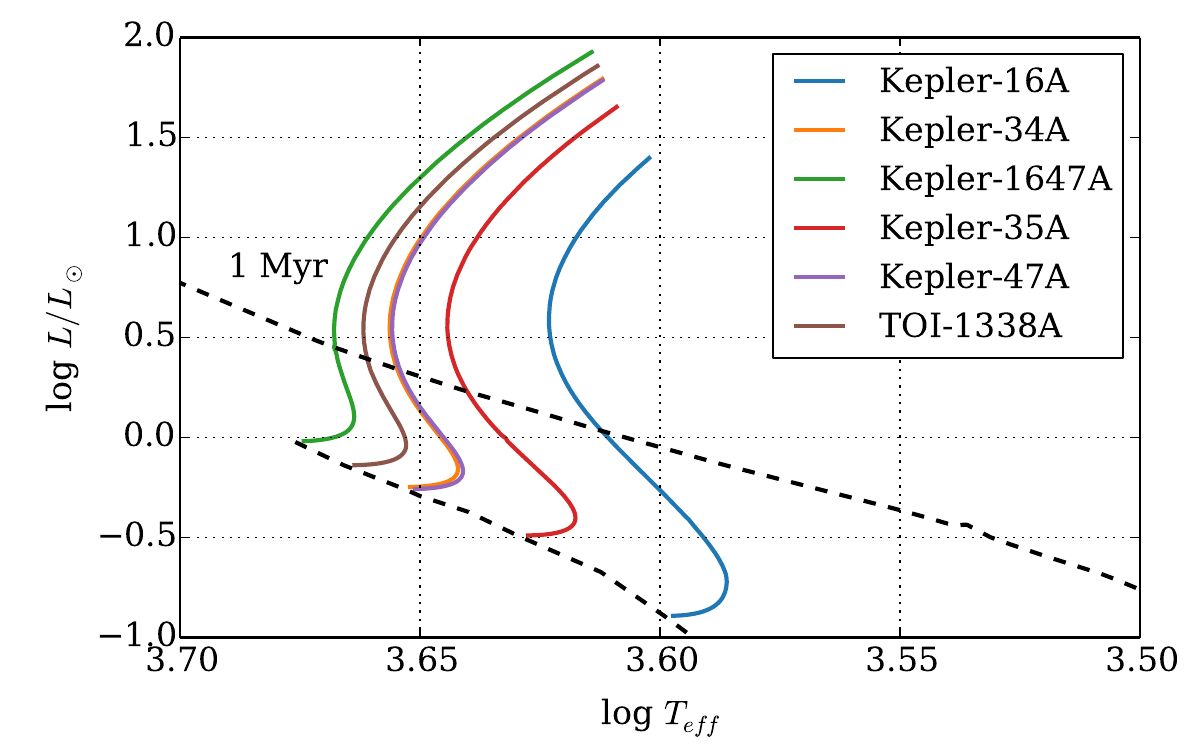}
\includegraphics[width=\columnwidth]{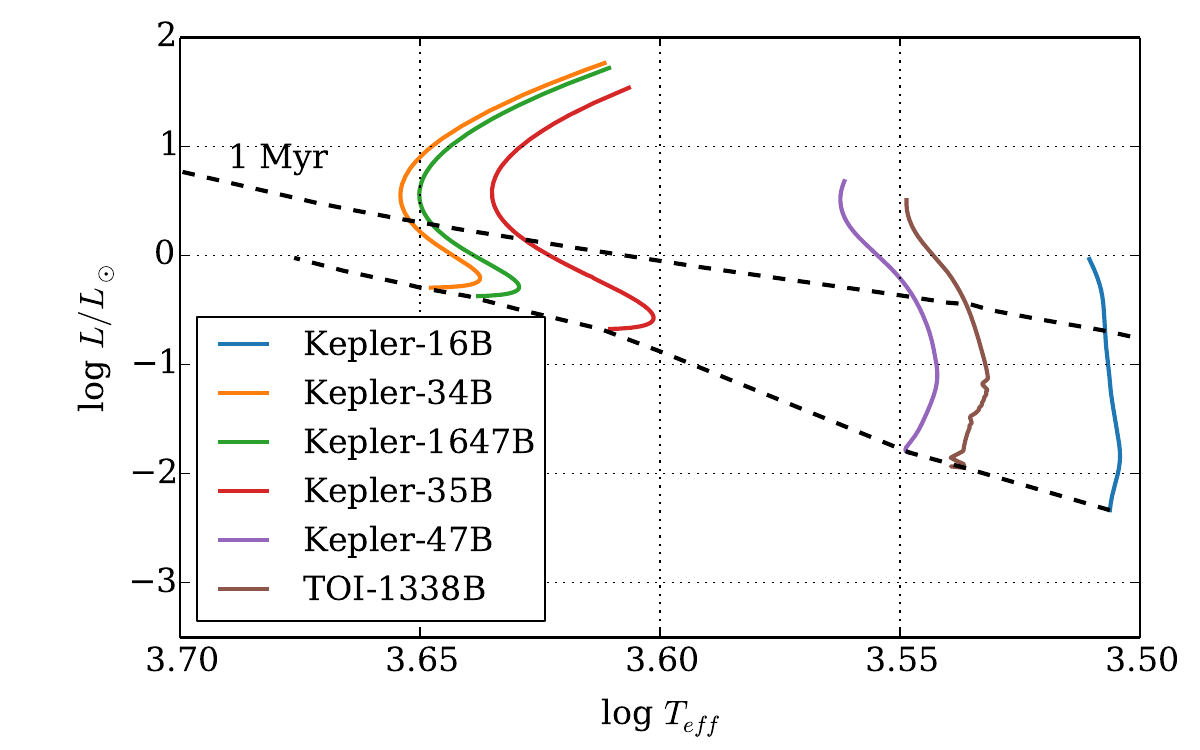}
\caption{Evolutionary tracks on the HR diagram for different circumbinary systems. The upper dashed line shows the stellar isochrone at 1 Myr, whereas the lower dashed line  corresponds to the zero-age main sequence (ZAMS).  Isochrones have been computed using MIST evolutionary models \citep{2016ApJ...823..102C}.}
\label{fig:HR}
\end{figure}

\begin{table*}
\caption{Binary stellar parameters for known circumbinary planet hosts taken from \cite{2021BAAS...53c1156G}. Systems considered in this work indicated by asterisks.  Left to right columns correspond to mass of the primary $M_1$, mass of the secondary $M_2$, radius of the primary $R_1$, radius of the secondary $R_2$, temperature of the primary $T_1$, temperature of the secondary $T_2$, binary luminosity $L_\star$, binary luminosity at 1 Myr, irradiation flux at the disc edge $F_{\rm in}$.   }              
\label{table1}      
\centering                                      
\begin{tabular}{c c c c c c c c c c}          
\hline\hline                        
System  &  $M_1(M_\odot)$ & $M_2(M_\odot)$ & $R_1(R_\odot)$ & $R_2(R_\odot)$ & $T_1({\rm K})$ & $T_2({\rm K})$ & $L_\star(L_\odot)$  & $L_\star {\rm at \;1 \;Myr} (L_\odot)$\  & $F_{\rm in} (L_\odot/{\rm AU}^2)$ \\
\hline
Kepler-16$^*$ & $0.69$ & $0.20$ & $0.65$ & $0.23$ & $4450$ & $3311$ & $0.15$ &1.2 & 25\\
Kepler-34 & $1.05$ & $1.02$ & $1.16$ & $1.09$ & $5913$ & $5867$ & $2.5$& 4.3 & 82 \\
Kepler-35$^*$ & $0.89$ & $0.81$ & $1.03$ & $0.79$ & $5606$ & $5202$ & $1.3$& 3.1& 100 \\
Kepler-38 & $0.95$ & $0.25$ & $1.76$ & $0.27$ & $5640$ & $3325$ & $2.1$ & 2.8 & 98\\
Kepler-47$^*$ & $1.04$ & $0.25$ & $1.76$ & $0.27$ & $5640$ & $3325$ & $0.8$& 2.5 & 400\\
Kepler-64 & $1.53$ & $0.38$ & $1.73$ & $0.41$ & $6407$ & $3561$ & $4.5$ & 4.5& 145 \\
Kepler-413 & $0.82$ & $0.54$ & $0.77$ & $0.48$ & $4700$ & $3463$ & $0.3$& 2.2& 213\\
Kepler-453 & $0.94$ & $0.19$ & $0.83$ & $0.21$ & $5527$ & $3226$ & $0.6$& 1.9& 55\\
Kepler-1647$^*$ & $1.21$ & $0.97$ & $1.79$ & $0.97$ & $6210$ & $5777$ & $5.1$& 5.1& 290\\
Kepler-1661 & $0.84$ & $0.26$ & $0.76$ & $0.28$ & $5100$ & $3585$ & $0.36$ & 1.8 & 52\\
TOI-1338$^*$ & $1.04$ & $0.3$ & $1.3$ & $0.3$ & $5990$ & $3316$ & $1.96$& 2.8 & 167\\
\hline                                             
\end{tabular}
\end{table*}

\begin{table}
\caption{Binary orbital parameters for all known circumbinary planet hosts.  Left to right columns correspond to the considered system, binary semi-major axis $\abin$, binary eccentricity $e_{\rm bin}$.  Data are taken from these papers: Kepler-16 \citep{2011Sci...333.1602D}; Kepler-34 \& 35 \citep{2012Natur.481..475W}; Kepler-38 \citep{2012Sci...337.1511O}; Kepler-47 \citep{2012ApJ...758...87O}; Kepler-64 \citep{2013ApJ...768..127S,2013ApJ...770...52K}; Kepler-413 \citep{2014ApJ...784...14K}; Kepler-453 \citep{2015ApJ...809...26W}; Kepler-164 \citep{2016ApJ...827...86K}; Kepler-1661 \citep{2020AJ....159...94S}; TOI-1338 \citep{2020AJ....159..253K}. }              
\label{table1}      
\centering                                      
\begin{tabular}{c c c}          
\hline\hline                        
System  &  $\abin$ (au)  & $e_{\rm bin}$  \\
\hline
Kepler-16 & $0.22$ & $0.16$ \\
Kepler-34 & $0.23$ & $0.54$  \\
Kepler-35 & $0.18$ & $0.14$ \\
Kepler-38 & $0.15$ & $0.1$ \\
Kepler-47 & $0.08$ & $0.02$ \\
Kepler-64 & $0.17$ & $0.21$ \\
Kepler-413 & $0.10$ & $0.03$ \\
Kepler-453 & $0.18$ & $0.05$ \\
Kepler-1647 & $0.13$ & $0.16$ \\
Kepler-1661 & $0.19$ & $0.11$ \\
TOI-1338 & $0.13$ & $0.16$ \\
\hline                                             
\end{tabular}
\end{table}

\begin{figure*}
\centering
\includegraphics[width=\textwidth]{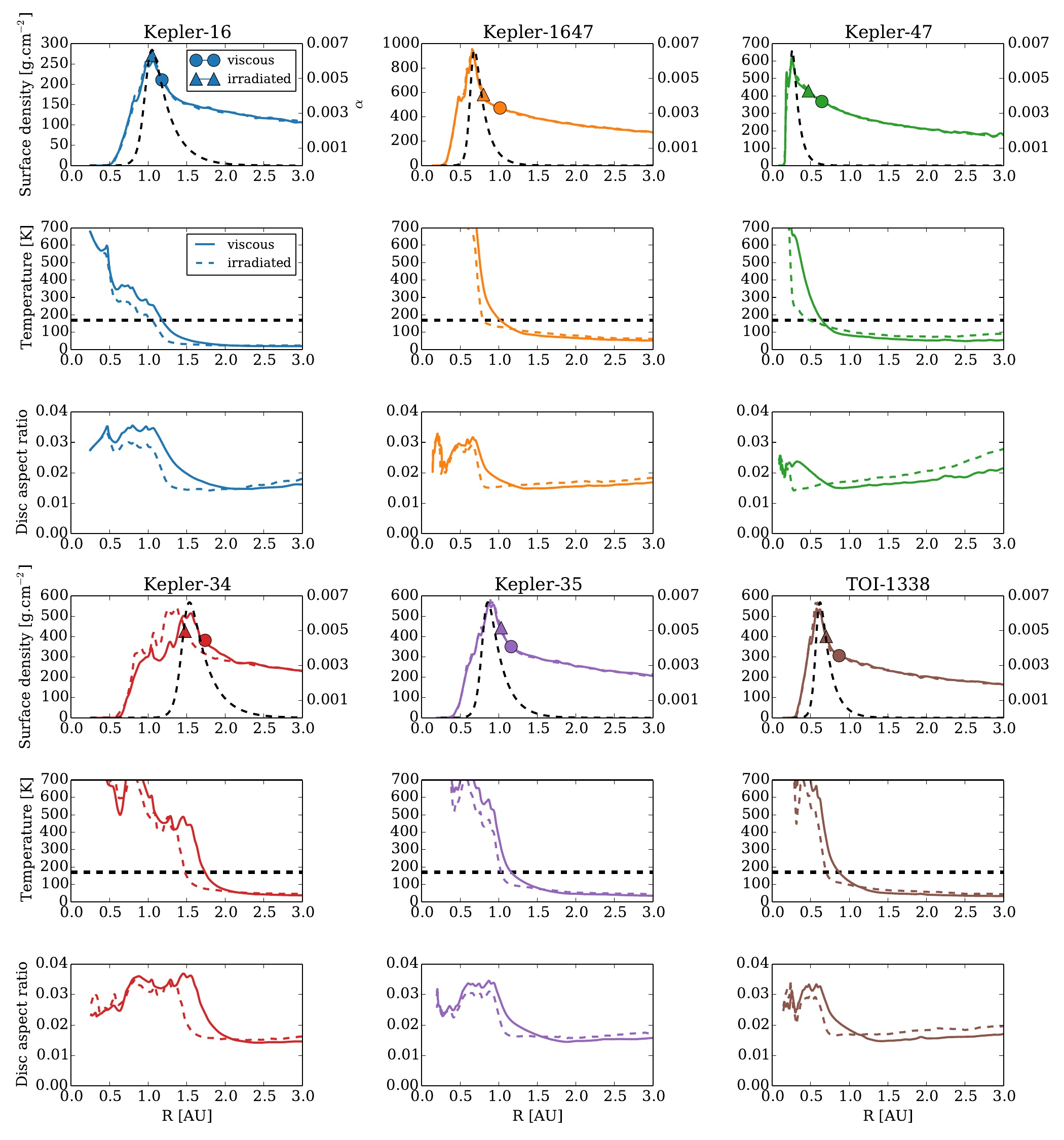}
\caption{Surface density, midplane temperature and aspect ratio profiles for the different circumbinary systems we considered. Solid lines correspond to models that include viscous heating whereas dashed lines are for models without viscous heating. All models include irradiation and shock heating. In the surface density profiles, the markers indicate the location of the ice line, which is also represented as an horizontal dashed line in the temperature plots. The black dashed lines in the surface density plots show the values of $\alpha$ in the viscously heated models - see the right hand vertical axis.}
\label{fig:results}
\end{figure*}

\begin{figure*}
\centering
\includegraphics[width=\textwidth]{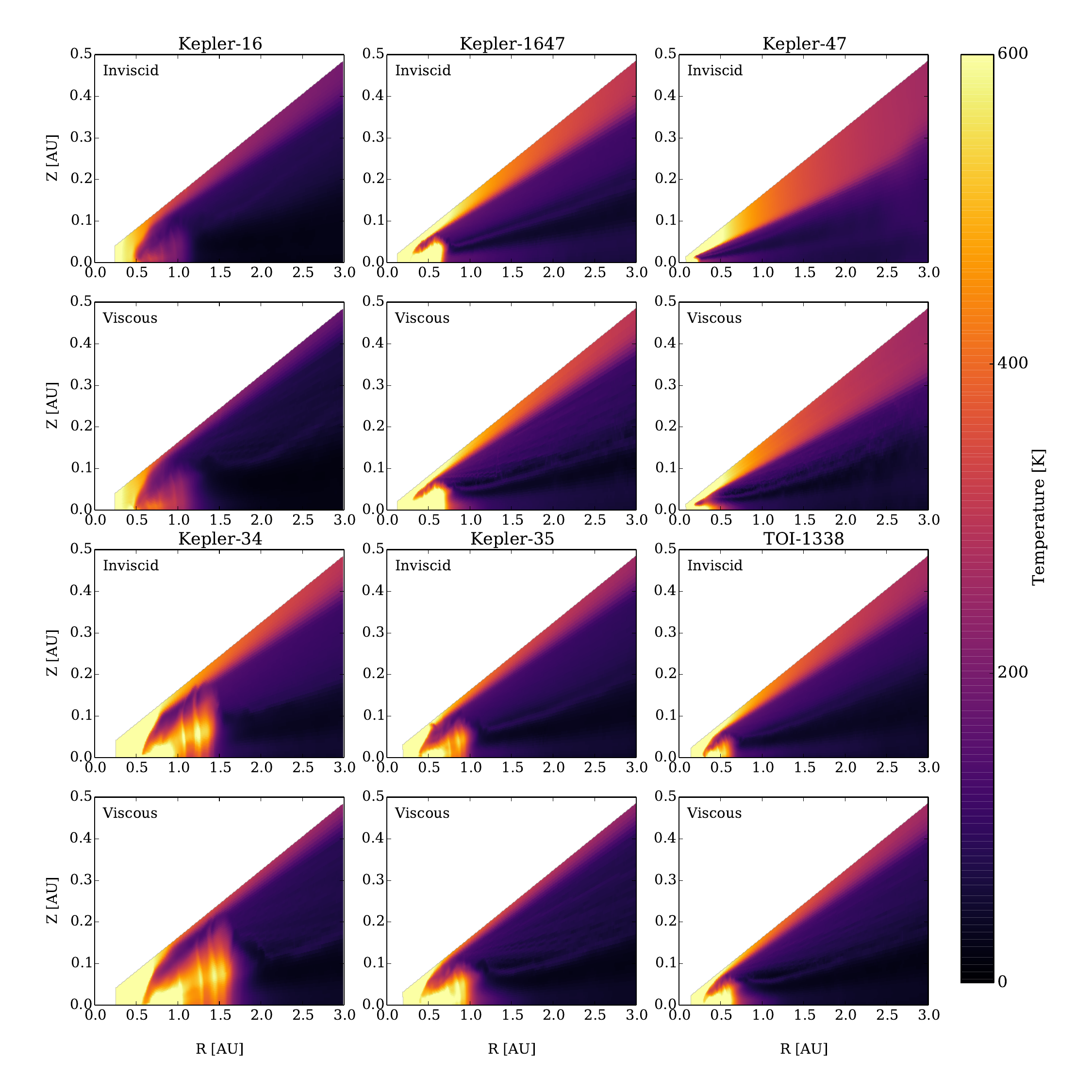}
\caption{Temperature distributions for each of the irradiated and viscously heated models that we considered.}
\label{fig:temp_maps}
\end{figure*}

\begin{figure}
\centering
\includegraphics[width=\columnwidth]{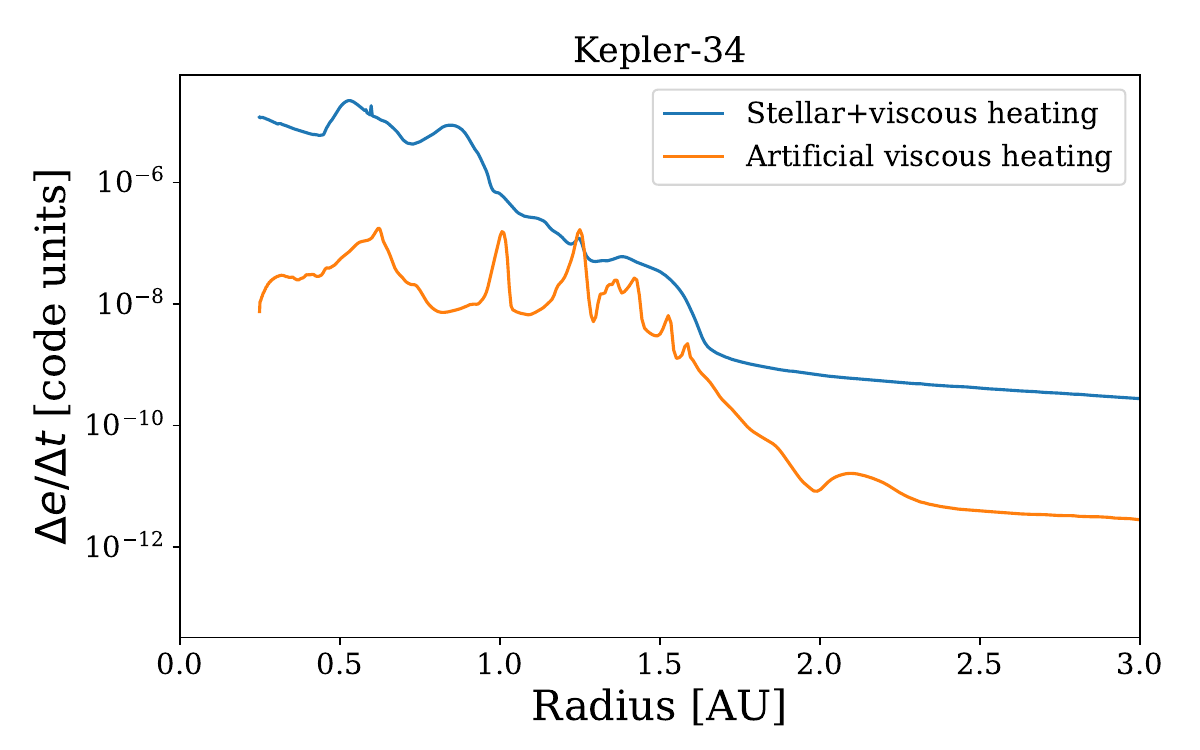}
\caption{Contribution of the different heating terms to the rate of specific internal change for Kepler-34.}
\label{fig:artificial}
\end{figure}

\begin{figure*}
\centering
\includegraphics[width=0.99\textwidth]{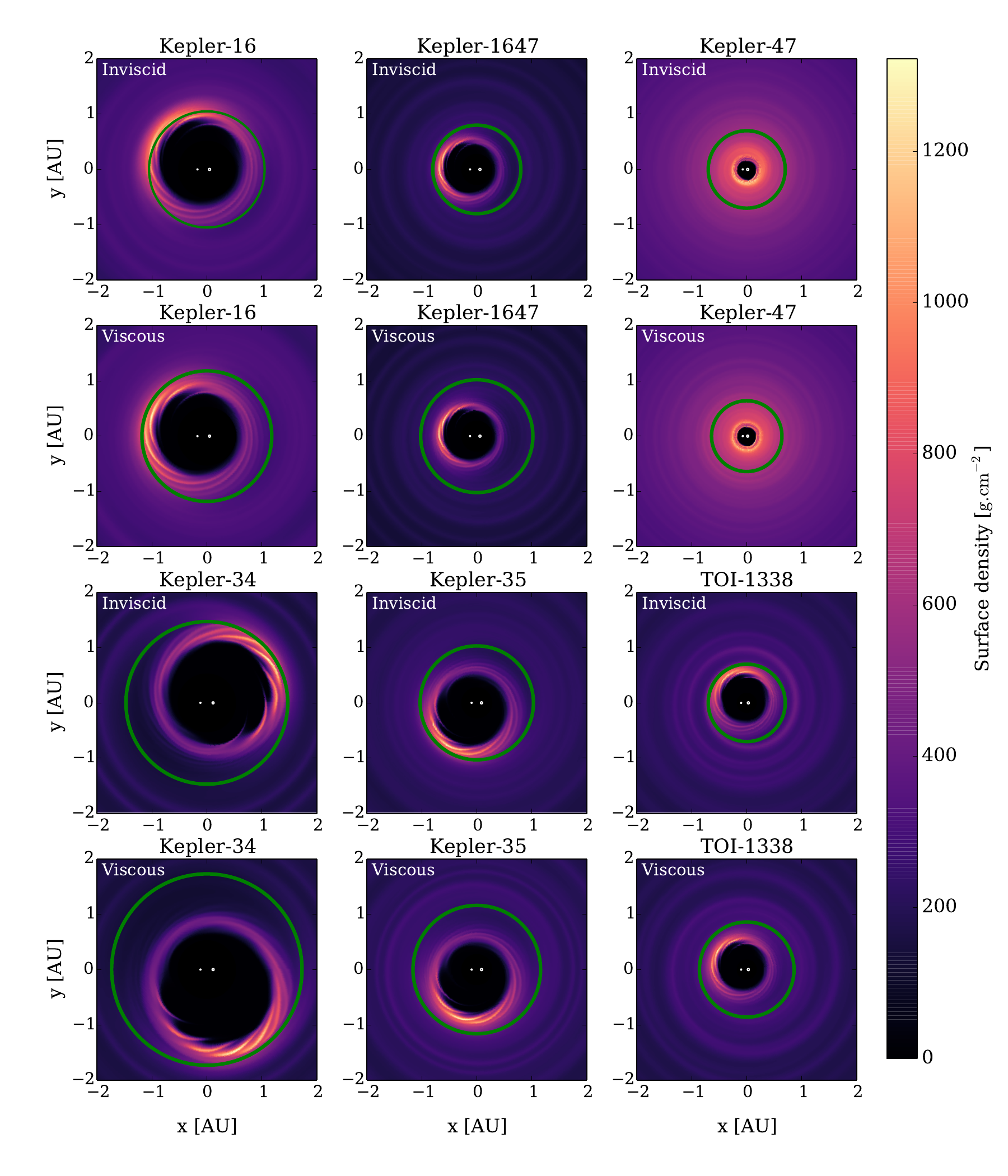}
\caption{Surface density distribution for each of the simulations. The location of the ice line is indicated by the green circle, while the locations of the stars are represented as white circles.}
\label{fig:densities}
\end{figure*}
\subsubsection{Grain properties and dust density profiles}

The chosen grain size distribution is composed of 10 size bins, logarithmically distributed between a minimum grain size of $a_{\rm min}=0.01$~$\mu$m and a maximum grain size of $a_{\rm max}=1$~mm.  The fraction of dust mass in each size bin is calculated assuming a dust-to-gas ratio $\epsilon=\Sigma_{\rm d}/\Sigma_{\rm g}=0.01$, where $\Sigma_{\rm g}$ and $\Sigma_{\rm d}$ are the gas and dust surface densities respectively. The grain sizes are described by a MRN distribution,  for which the number density of grains in the range $[a,a+da]$ is given by $n(a) da \propto a^{-3.5}$ \citep{1977ApJ...217..425M}. We note that the  RADMC-3D calculations were performed using these 10 grain sizes as 10 different dust species, each having its own opacity. The opacity of each grain size  is calculated using optool \citep{2021ascl.soft04010D} and by considering the DSHARP opacities \citep{2018ApJ...869L..45B}. These assume that the  dust is composed of water ice, astronomical silicates, troilite, and refractory organic material, with volume fractions being 36\%, 17\%, 3\%, and 44\%, respectively. This results in a  bulk density of the dust mixture of $\rho_{p} = 1.675 {\rm g}\cdot {\rm cm}^{-3}$. 
The dust volume density is defined by imposing a Gaussian vertical profile:
\begin{equation}
\rho_{\rm d}(R,z,a)=\frac{\Sigma_{\rm d}(R,a)}{\sqrt{2\pi}H_{\rm d}(R,a)}\exp\left(\frac{-z^2}{2H_{\rm d}(R,a)^2}\right).
\label{eq:rhodust}
\end{equation} 
To account for dust settling, the particle scale height $H_{\rm d}(R,a)$ varies with grain size $a$ and is given by:
\begin{equation}
H_{\rm d}(R,a)=H_{\rm g}(R)\sqrt{\frac{\alpha}{\st+\alpha}},
\end{equation} 
where $\st$ is the Stokes number calculated in the Epstein regime:
 \begin{equation}
 \st=\frac{\rho_p a}{\rho c_s} \Omega,
 \end{equation}
 For low levels of turbulence, particles tend to be strongly confined in the disc midplane, resulting in a very optically thick region. To avoid this, we employ a residual viscosity with $\alpha_{\rm res}=3\times 10^{-5}$ that limits the scale height of  mm-sized grains to values of $H_{\rm d}=0.1-0.2 H_{\rm g}$, consistent with the values reported in \citep{2021MNRAS.508.4806P} (see their Fig.~6).

\section{Model setup}
\subsection{Initial conditions}

The initial gas surface density profile  is given by:
\begin{equation}
\Sigma_{\rm g}=f_{\rm gap}\, \Sigma_{{\rm g},0}\left(\frac{R}{R_0}\right)^{-1/2},
\label{eq:sigma0}
\end{equation}
where  $\Sigma_0$ is the surface density at $R_0=\abin$ and which is defined such that the mass of the gas disc is $5\%$ of the total binary mass within 40~au. $f_{\rm gap}$ is a gap-function used to initiate the disc with an inner cavity (assumed to be created by the binary), and is given by:
\begin{equation}
f_{\rm gap}=\left(1+\exp\left[-\frac{R-R_{\rm gap}}{0.1R_{\rm gap}}\right]\right)^{-1},
\end{equation}
 where $R_{\rm gap}$  is the  estimated gap size which is set to $R_{\rm gap}=2.5$ $\abin$. The initial azimuthal velocity is set to the Keplerian velocity, whereas both the initial radial and meridional velocities are set to zero.

The initial temperature in the disc corresponds to the stellar irradiation-dominated temperature:
\begin{equation}
T_{\rm irr}(R)=\left(\frac{L_\star}{4\pi R^2 \sigma_{\rm SB}}\right)^{1/4},
\label{eq:T0}
\end{equation} 
where $L_\star$ is the total stellar luminosity. Using Eqs. \ref{eq:sigma0} and \ref{eq:T0}, one can deduce the 3-dimensional gas density which, combined with the dust volume density given by Eq. \ref{eq:rhodust}, can be subsequently employed in RADMC-3D  to compute the dust temperature.  The new gas temperature is then  simply obtained by considering the gas and dust temperatures to be equal, which is a reasonable assumption for $\mu$m-sized  grains that tend to be strongly coupled to the gas. After this radiative transfer calculation, the new gas density is then calculated by FARGO3D and we iterate this procedure every binary orbit until we converge to a quasi-equilibrium solution, which is typically obtained after $\sim 5$ iterations. Over longer evolution timescales, we increase the interval between two radiative transfer calculations to $10$ and then $100$ binary orbits until a quasi-stationary state is reached.

\subsection{Binary parameters}

It can be reasonably  expected that over the $\sim 10^7$~years of circumbinary disc evolution, the characteristics of the central binary (temperature, luminosity...) change significantly as well. In the context of circumbinary planet formation, this is an important issue as it may influence the amount of stellar heating received by the disc, resulting in a change in the structure of the inner cavity.  This is illustrated in Fig.~\ref{fig:HR} where we show evolutionary tracks on a HR diagram for the circumbinary systems that we will focus on in this paper. The stellar luminosity continuously  decreases as it follows the Hayashi track,  until it reaches a value close to the luminosity at the zero-age main sequence. The overplotted stellar isochrone at 1 Myr confirms that at the formation epoch of a circumbinary planet, the binary star's combined luminosity was probably much higher than it is at the present day. For the $11$ known circumbinary planet systems, the values for the luminosities after 1~Myr of evolution are listed in Table 1, and the orbital parameters are listed in Table 2.  Representing the binary as a point source located at the centre-of-mass, the last column gives the corresponding irradiation flux $F_{\rm in}$ at a distance of $\abin$. In terms of the values for $F_{\rm in}$, the circumbinary systems that are presented in Fig.~\ref{fig:HR} span the range of values that are obtained across the whole set of known circumbinary systems. Hence, we focus on these six particular systems in the rest of this paper. 

 We note that over an evolution timescale of 1~Myr, the binary separation is also expected to be impacted by its interaction with the circumbinary disc \citep[e.g.][]{2024MNRAS.tmp..195T}. For simplicity, we here assume a fixed value for $\abin$ but it should be kept in mind that compared to its present value, the binary semi-major axis was also probably slightly larger when circumbinary planets formed.

\section{Results}
\label{sec:results}
For each system considered, Fig.~\ref{fig:results} shows the azimuthally averaged surface density, midplane temperature and disc aspect ratio profiles at time=$10^3\,\Tbin$.  This corresponds to the evolution timescale over which the circumbinary disc structure reached a quasi-steady state in 3-dimensional models \citep{2023A&A...670A.112P}. We see that inviscid discs and those with viscous heating have similar profiles in all quantities (especially the surface density), except for Kepler-34 where  the size of the cavity in the model with viscous heating is slightly larger. As discussed below, this largely arises because the viscous dissipation is localised to the regions just outside of the inner cavity, but is also because the localised viscous heating rate is similar to the shock dissipation and irradiation heating rates, so that viscosity does not play a dominant role in determining the density and temperature structure of the inner regions of circumbinary discs. 

The filled circles and triangles in Fig.~\ref{fig:results}, representing the locations of the snow lines at the disc  midplanes, show that these are located close to the cavity edge at distances of between $\sim 0.7$-1.8~au from the central binary, depending on which system is being considered and whether or not viscous heating has been included. As expected, the effect of including viscous heating is to move the snow line outwards, but in general this is a small effect. In the models with viscous heating, the snow line sits out beyond the location where the viscosity is at its maximum value. Hence, it is not just the local viscous heating that plays a role but also the radiative diffusion radially outwards of the viscously generated heat that is important.\footnote{We note here that radiative diffusion of spiral wave heating is not included in our models, and will likely lead to a slight increase in the location of the snow line. This will be investigated in future work.}


Looking at the $H/R$ profiles in Fig.~\ref{fig:results}, we observe slightly larger disc aspect ratios in the innermost regions of the discs with viscous heating included.  In all models, the disc aspect ratio exhibits significant bumps near the inner cavity where all the heating processes operate most effectively. This leads to the formation of a self-shadowed region in the outer disc. Compared to an inviscid disc, a viscous disc tends to have a  higher aspect ratio in the vicinity of the density peak because viscous heating scales with gas density and the viscous dissipation is concentrated in this region.  The resulting enhanced self-shadowing effect leads to a slightly smaller temperature and aspect ratio in the outer disc in comparison to an inviscid disc. 

In Fig.~\ref{fig:temp_maps}, we show the azimuthally-averaged temperature distributions in the $(R-Z)$ plane for each system considered.  Again, only small differences between models with/without viscous heating are observed.  Compared to a viscous disc, the regions beyond the inner cavity and density maximum of an inviscid disc tend to have a slightly higher temperature at the surface, presumably because more stellar radiation is intercepted in the inner disc regions in the viscous models because of the higher aspect ratios there. 

We also observe that in the vicinity of the density peak and out just beyond it the temperature of an inviscid disc can reach $T\sim 500-600$~K in the midplane. This is a consequence of the shock dissipation of density waves excited by the binary, which provides an important source of additional heating, as can be seen in the contours in Fig.~\ref{fig:temp_maps}, where the wave-like structures illustrate the heating contribution from shock heating. The effect is particularly pronounced for the Kepler-34 run, 
and Fig.~\ref{fig:artificial}, which shows the different contributions to the change in specific internal energy, shows that the heating efficiency due to shocks can be of the same order as the viscous and stellar heating combined near the tidally-truncated cavity.

The importance of the dissipation of spiral shocks in setting the disc thermodynamical structure of circumbinary discs has also been emphasised by \cite{2016ApJ...816...94V}. In their models, which involve globally viscous discs in which the torque from the central binary prevents inwards mass accretion at the inner edge of the cavity, and in which heating is provided by viscosity, shock dissipation and stellar irradiation, the snow line was typically found to lie beyond 3-6~au, depending on the model and the evolutionary stage under consideration. These values are clearly much larger than those obtained by the models presented here, and arise because of the fundamentally different underlying disc models that they considered.


The locations of the snow lines for the different circumbinary systems considered here are shown in Fig.~\ref{fig:densities} by the green circles over-plotted on the surface density images.  As described above, the snow line is located slightly further away from the binary in viscous compared to inviscid models, typically at radial distances $\lesssim 1.5$~au. More importantly, we find that in both inviscid and viscous discs, the snow line tends to be located close to the density maximum and inner cavity.  Hence, there is only a narrow range of radial locations in the discs where the midplane temperature  is $\gtrsim 170$~K, such that it is difficult to envisage rocky planets that contain no icy material being able to form. This statement is reinforced by the fact that high levels of hydrodynamical turbulence are expected to be present near the inner cavity, such that pebble accretion onto seed protoplanets would proceed in the inefficient, 3D regime in which the pebble layer has a scale height that exceeds the Hill sphere radius of the accreting body.  In situ grain growth is also expected to be difficult in this region due to the high collision velocities that are also expected to arise because of the turbulence. Indeed, \citep{2021MNRAS.508.4806P} found that at distances $\lesssim 8-9 \, \abin$, which, for example, would correspond to $\sim 2$~au for Kepler-16 and $\sim 0.7$~au for Kepler-47, collision velocities are higher than the fragmentation velocities of silicate aggregates. Taken together, these factors suggest that planets forming in circumbinary discs are likely to form exterior to the snow line, and hence will be composed of a significant water ice fraction. 

\section{Discussions and conclusions}
In this paper, we have presented the results of 3-dimensional hydrodynamical simulations that are tightly coupled to Monte Carlo radiative transfer calculations to determine the location of the snow line in circumbinary discs. We choose binary parameters that match those of binary systems that are known to host circumbinary planets (e.g. Kepler-16, -34). 

The simulations include a cooling scheme where the disc temperature is relaxed towards an irradiation temperature on a finite cooling timescale that was calculated using realistic opacities. The irradiation temperature is determined by performing Monte Carlo radiative transfer calculations using RADMC-3D, in which localised viscous heating due to hydrodynamical turbulence was incorporated as an additional heating source for a subset of the models. Our method involves evolving the hydrodynamical and thermodynamical states of the disc in lockstep by performing radiative transfer calculations at different stages during the hydrodynamical simulations. Hence, the evolution of the density and temperature structures of the disc feed back on each other as a quasi-steady state is established, and the effects of stellar irradiation, shock heating due to the dissipation of binary-induced spiral density waves, and viscous dissipation (when switched on) are included in the models.

We find that the snow line lies close to the density maximum that sits just outside of the tidally-truncated cavity in all the circumbinary discs. In discs that include viscous heating, the ice line lies further out but only by a small amount. The similar disc structures obtained in viscous and inviscid models are a consequence of shock heating, stellar irradiation and viscous dissipation all providing a similar amount of heating in the disc, and because these heat sources are quite localised in the inner regions of the disc close to the cavity. 

In the outer disc,  heating at shocks and through viscous dissipation are obviously much smaller. The latter depends on the level of turbulence operating in the disc and how this is spatially distributed. The origin of the turbulence is a parametric instability that only operates close to the inner cavity where the disc becomes visibly eccentric, and hence viscous heating is only important in a narrow region close to the surface density maximum. This results in typical snow line locations of  $\lesssim1.5$~au from the binary. At these distances, pebble accretion is believed to be inefficient due to the vertical stirring of dust and pebbles induced by the turbulence. Similarly, the high relative collision velocities there reduce the likelihood of in-situ grain growth to sizes sufficient to trigger the streaming instability.  

These considerations indicate that it will be difficult to form rocky planets that are devoid of significant fractions of water ice in circumbinary discs, and we suggesting that circumbinary planets should preferentially be icy and not rocky.  A similar conclusion was reached by \cite{2013ApJ...768L..15C} using a simpler disc model than the one presented in this paper. This statement should be particularly true for planets evolving on orbits that are coplanar to the binary orbit.  This is the case for most of the circumbinary planets detected so far but we note that this may be partly due to an observational bias, as misaligned circumbinary planets are  more difficult to detect due to their non-periodic transits \citep{2014A&A...570A..91M}. Nevertheless, the observation of misaligned circumbinary discs with ALMA \citep[e.g.][]{2019ApJ...883...22C}, and the detection of  Kepler-413b \citep{2014ApJ...784...14K} and Kepler-453b \citep{2015ApJ...809...26W} whose orbital planes are slightly misaligned with respect to the binary orbital plane, indicate that misaligned circumbinary planets do exist.  It has even been proposed that circumbinary planets around polar orbits may form with comparable efficiency to coplanar planets (\cite{2021ApJ...920L...8C}). Such systems may form in circumbinary discs that are polar aligned to the binary orbit. For an eccentric  binary, a circumbinary disc with  sufficient initial inclination with respect to the binary orbit may achieve a polar configuration due to viscous dissipation (\cite{2017ApJ...835L..28M}). Compared to the coplanar case, however, the tidal torque exerted on a disc in a polar state is much weaker, resulting in an inner disc edge that lies closer to the binary \citep{2018MNRAS.479.1297M,2019ApJ...880L..18F}, plausibly inside the ice line. This suggests that if rocky planets can form in circumbinary discs, these would preferentially be found on orbits that are highly inclined with respect to the binary orbit. 

The prediction that circumbinary planets should be icy not rocky receives tentative support from current observations of circumbinary planets (CBPs). All CBPs discovered so far have radii $\ge 3$~R$_{\oplus}$, and appear to have significant hydrogen-helium envelopes, so in this regard they are more similar to the ice and gas giant planets of the Solar System than the rocky terrestrial planets. Furthermore, the known population of CBPs differs significantly compared to the exoplanet population that orbits around single stars which is dominated by smaller super-Earths and mini-Neptunes. 

Attempts have been made to search for additional transiting CBPs orbiting eclipsing binaries in the Kepler data, using customised algorithms that account for the large transit timing variations inherent in such systems \citep{2021AJ....162...84M, 2022BAAS...54e2702M}, but these searches have failed to find any new planets even though they should have been sensitive to finding planets smaller than 3~R$_{\oplus}$ if they were present in the data \citep{2022BAAS...54e2702M}. 

While it is currently not possible to determine the compositions of the cores of the known CBPs, it is tempting to infer that the lack of smaller CBPs arises because their formation involves incorporation of a substantial fraction of icy material that augments the masses of their cores, leading to the accretion of significant hydrogen envelopes. Future observations by TESS, and discoveries by the PLATO mission, will enhance the population of known circumbinary planets and will provide direct measurements of their radii. Radial-velocity surveys such as the Binaries Escorted By Orbital Planets (BEBOP; Martin et al. 2019) survey are also detecting and discovering CBPs using the radial-velocity method \citep{2022MNRAS.511.3561T,2023NatAs...7..702S}. Hence, it is to be hoped that high-precision radial-velocity measurements of future planets discovered by transit surveys will provide measurements of their density, and by extension constraints on their bulk compositions. Eventually, this will allow a CBP mass-radius diagram to be constructed, providing insight into the compositional diversity of the circumbinary planet population.

\section*{Acknowledgments}
Computer time for this study was provided by the computing facilities MCIA (M\'esocentre de Calcul Intensif Aquitain) of the Universite de Bordeaux and by HPC resources of Cines under the allocation A0110406957 made by GENCI (Grand Equipement National de Calcul Intensif). RPN acknowledges support from the Leverhulme Trust through grant number RPG-2018-418 and from STFC through grants ST/X000931/1 and ST/T000341/1.

\end{document}